\documentclass[aps,pra]{revtex4}

\def\duzomniejsze{<\kern-.7mm<}
\def\duzowieksze{>\kern-.7mm>}

\def\textbf#1{{\bf #1}}
\def\beq{\begin{equation}}
\def\eeq{\end{equation}}
\def\be{\begin{equation}}
\def\ee{\end{equation}}
\def\ben{\begin{eqnarray}}
\def\een{\end{eqnarray}}
\def\beqa{\begin{eqnarray}}
\def\eeqa{\end{eqnarray}}
\def\eea{\end{array}}
\def\bea{\begin{array}}
\newcommand{\bei}{\begin{itemize}}
\newcommand{\eei}{\end{itemize}}
\def\ra{\rangle}
\def\la{\langle}

\def\pcal{{\cal P}}

\def\dt#1{{{\kern -.0mm\rm d}}#1\,}
\def\ypodpis{\raise4mm\hbox{$\omega$}}

\def \ILO {I_{LO}} 
\def \W {I_{c}}
\def \WP {I_{LOCC}(\pcal)}
\def \WL {E_D(\pcal)}
\def \HL {H_B(\pcal)}
\def \HP {H_{LOCC(\pcal)}}

\def \Icor {I_{cor}(\pcal,\rab)}
\def \s {\,\,\,\,}
\def\rab{\rho_{AB}}
\def\D{\Delta_q}
\def\Dc{\Delta_c}
\def\r{\rho}
\def\sza{\sigma_z^{(A)}}
\def\szb{\sigma_z^{(B)}}
\def\sxa{\sigma_x^{(A)}}
\def\sxb{\sigma_x^{(B)}}

\def\syb{\sigma_y^{(B)}}
\def\Sx{\Sigma_x}
\def\Sz{\Sigma_z}
\def\Sxl{\Sigma_x^{LOCC}}
\def\Szl{\Sigma_z^{LOCC}}
\def\M{M}
\def\N{N}
\def\Ml{M_{LOCC}}
\def\Nl{N_{LOCC}}

\begin{document}
\title{A new type of complementarity between quantum and classical
information}
\author{Jonathan Oppenheim$^{(1)(2)}$, Karol Horodecki$^{(3)}$, Micha\l{} Horodecki$^{(2)}$, Pawe\l{}
Horodecki$^{(4)}$ and Ryszard Horodecki$^{(2)}$}

\affiliation{$^{(1)}$
Racah Institute of Theoretical Physics, Hebrew University of Jerusalem, Givat Ram, Jerusalem 91904, Israel}
\affiliation{$^{(2)}$Institute of Theoretical Physics and Astrophysics,
University of Gda\'nsk, Poland}
\affiliation{$^{(3)}$Faculty of Mathematics, University of Gda\'nsk, Poland}
\affiliation{$^{(4)}$Faculty of Applied Physics and Mathematics,
Technical University of Gda\'nsk, 80--952 Gda\'nsk, Poland}

\begin{abstract}
  Physical systems contain information which can be divided between
  classical and quantum information.  Classical information is locally
  accessible and allows one to perform tasks such as physical work, 
  while quantum information allows one to perform tasks
  such as teleportation.  It is shown that
  these two kinds of information are complementarity in the sense that two
  parties can either gain access to the quantum information, or to the classical information
  but not both.  This complementarity has a form very similar to the 
  complementarities usually encountered in quantum mechanics.   For
  pure states, the entanglement plays the role of Planck's constant.
  We also find another class of complementarity relations which applies to
  operators, and is induced when two parties can only perform local operations
  and communicate classical. In order to formalize this notion we define the
  restricted commutator.  Observables such as the parity and phase of two
  qubits commute, but their restricted commutator is non-zero.  It is also found
  that complementarity is pure in the sense that can be ''decoupled'' 
  from the uncertainty principle.
\end{abstract}
\maketitle

\section{introduction}
The idea of complementarity appeared together with the birth of quantum
mechanics in 1926 \cite{Pauli}
\cite{Heisen,Bohr}. In quantum mechanics, observables 
such as the position and momentum of a
particle are complementary.  An accurate measurement of momentum will
make a subsequent measurement of position yield random
results - the position information is destroyed during the momentum measurement.
In this work, we find new complementarities which arise in the context of
quantum information theory.

In quantum information theory, one is often interested in
measurements that can be made by two parties in distant labs
who can only communicate classically and perform local operations (LOCC).  We
find that this restriction on the class of allowable operations
induces new complementarity principles.  In this paper, we will
explore two different types of complementarity.

The first, is a complementarity relationship
between classical and quantum information.  
This is a generalization
of an idea we briefly noted in \cite{JHHH}.  In
Section \ref{sec:clasandquant} we will first demonstrate how to divide
information into classical and quantum parts in an operational way.  
The essentially idea is that classical information is locally accessible
and can be used to perform tasks such as physical work, while quantum 
information can be used to perform tasks such as teleportation and double
dense coding.  Classical information can be measured by how many pure
separable bits can be obtained from a state, while one can obtain quantum
information by distilling singlets from the state.

Then in Section
\ref{sec:comp} we show that the two types of information are complimentary.
One finds that one
can exploit the quantum information to perform teleportations, but then the
ability to use the classical information is completely destroyed.  Or, one can
obtain classical information (locally accessible), but then the ability to perform teleportations
is completely destroyed.  We will show that this idea can be expressed as an
information theoretic bound which has the same form as an uncertainty relation.
For pure states, the bound has the feature that the entanglement plays the role
of Planck's constant $\hbar$.  Pure states can also be thought of as the 
counterpart of coherent states (i.e. minimum uncertainty wave packets).

In Section \ref{sec:loccom} we introduce a complementarity principle involving
individual measurements.  For example, if one has two qubits, one can globally
measure the parity and phase.  However, when two parties each hold one of
the qubits in distant labs, then they find that all parity and phase
measurements will be complimentary.  They can measure the parity of the
state, or the phase of the state, but not both.  To quantify this, we introduce
the idea of the LOCC commutator.  We find that two observables can be
complimentary without being uncertain, demonstrating that the two
concepts can be decoupled.
We conclude in Section \ref{sec:conc}, and mention some open questions.

\section{classical and quantum information}
\label{sec:clasandquant}
Before showing how information can be divided into classical and
quantum contributions, a few preliminaries are necessary.
Consider a bipartite state $\r_{AB}$ composed of $n$ qubits which could be
shared between two parties Alice and Bob \cite{f1}.  
Let $\r_A$ and $\r_B$ be the
reduced density matrix for each party, and let $n_A$ and $n_B$ be the number of
qubits that each party holds.

The total information in the state is given by 
\beq
I=n-S(\r_{AB}) 
\label{eq:workinfo}
\eeq
where the
symbol $S(\r)$ is the von-Neumann entropy of a state $\r$. The more  we know
about the state of a system, the lower it's entropy and the greater the
informational content of the state.
This information can
be divided into local information contents $I_A=n_A-S(\r_A)$, $I_B=n_B-S(\r_B)$
and the mutual information $I_M=S(\r_A)+S(\r_B)-S(\rab)$ so that
\be
I=I_A + I_B + I_M
\label{eq:breakdown}
\ee
Typically, the mutual information $I_M$ is used as a measure of the total correlations
between $\r_A$ and $\r_B$.  It tells us 
how much information the two systems have in common.


In classical information theory \cite{thomascover}, a classical
system \cite{f2} $\r_{cl}$ has a mutual
information which is always smaller than the total Shannon entropy H of the
state. \be
I_M(\r_{cl})\leq H(\r_{cl})
\label{eq-mutual-classical}
\ee
This means that
the correlations are always accompanied by a lack of information
about the total system: only mixed states can have nonzero correlations.
Also, for two $d$-level classical systems,  the correlations cannot
exceed $\log d$.

For quantum system there is no restriction like
(\ref{eq-mutual-classical}). Therefore pure states can
contain correlations, and the mutual information can be twice as much as
in the classical case. In general we have
\be
I_M(\r_{qu})\leq 2\log d
\ee
Thus two qubits can share two bits of mutual information as in the case of a
maximally entangled state such as the singlet
\be
\psi^-={1\over \sqrt 2} (|01\ra-|10\ra) \s .
\label{eq:singlet}
\ee

There is a basic question: For the singlet, what is the meaning of the fact
that the amount of mutual information is {\it two}? One possible answer comes
from superdense coding \cite{Bennett92}. By using a singlet, one can
communicate 2 bits of information through 1 qubit \cite{BSST-entass}.
It has also
been argued that the additional correlations are related to negative conditional
entropies \cite{cerf}. We will propose a different answer to this
fundamental question. We will argue that $2$ is not equal to $1+1$ but rather it
is equal to {\it either $1$ or $1$}.  In other words, the two bits of mutual
information can be divided into one bit of quantum information, and one bit of
classical information, but that these two types of information are complimentary
- one can retrieve the one bit of classical information, or the one bit of
quantum information, but not both.  In general, we shall find that the
correlations of a quantum state consist of two {\it complementary} parts - one
classical, the other quantum.

We now show how to divide the total information into quantum
and classical parts.
The first attempt at quantifying quantum contents of
correlations other than through entanglement is
due to Zurek \cite{Zurek}. Quantifying classical correlations of a
quantum state and the division into classical and quantum
correlations was proposed in \cite{HV}.
An operational proposal of quantifying different types of
correlations was first proposed in \cite{JHHH}.  Another method,
using the entanglement of purification was given in \cite{THDL-entpur}.
Here we will follow \cite{JHHH}
where the division emerges from thermodynamical considerations.

The idea is to define information operationally - {\it classical information}
is defined to be information that can be manipulated into a locally accessible form.  In other
words, it is the information that can be localized by two parties, Alice
and Bob, and used to perform classical tasks.  For example,
as discussed in \cite{JHHH}, it can be used to extract real physical work from a local
heat bath, using a Szilard engine \cite{szilard} or (for quantum states)
a Von Neumann engine \cite{vn}.
Information that is locally accessible is equal to the maximum amount of
work which can be drawn from a local
heat bath by Alice and Bob under LOCC in units of $kT$ where
$T$ is the temperature of the bath and $k$ is Boltzmann's constant.
We will henceforth set $kT=1$ so that the amount of work drawn
is measured
in bits.  One can think of the classical information as the maximum amount
of pure separable states which can be extracted from a state.
We will therefore talk of {\it extracting} classical information from
a state, with the understanding that it could refer to extracting
physical work from the information encoded in the state, or extracting
a number of pure separable states.
It should also be mentioned that this definition of classical
information is independent of the
interpretation of quantum mechanics one uses
(Copenhagen, Many Worlds, Bohmian, etc.).

On the other hand {\it quantum information} is defined to be the information which can be
used to perform tasks which have no classical counterpart, such as teleportation
and double-dense coding.  One bit of quantum information, can be used
to teleport one qubit.
One can think of teleportation (sending qubits) as analogous to a form of
quantum logical work \cite{balance}.

Let us first look at that case where
Alice and Bob are only allowed to perform local operations (LO).
In this case, the amount of information $I_{LO}$ they can obtain
is
\beqa
I_{LO}&=&n_A-S(\rho_A)+n_B-S(\rho_B)\nonumber\\
&=&I_A + I_B
\label{eq:wlo}
\eeqa


On the other hand, if we allow Alice and Bob to perform any local operations (LO)
and send qubits to each other through a classical channel (CC) \cite{f3}, then they
will be able to obtain more information from the state by exploiting
correlations.
Alice and Bob can then transform the state $\rab$ into a new state $\rab'$
such that the amount of local information is maximized.  The amount
of obtainable local information is
\beqa
\W &=& I_A(\r_A')+I_B(\r_B')\nonumber\\
&=& n-S(\r_A')-S(\r_B')
\label{eq:wlocc}
\eeqa

then the difference
\beq
\Delta_{c} \equiv \W -\ILO
\label{eq:deltac}
\eeq
tells us the additional information that can be obtained, if the two parties
are also given access to a classical communication channel (CC).
%
Since the channel
is classical, we will refer to $\Delta_c$ as the {\it classical deficit}.




On the other hand, if Alice and Bob had access to a quantum channel (QC) rather
than to the classical channel,
they will be able to localize all the information (and draw all the work from the state).
The information they can obtain under these operations (LOQC) is just
\beqa
I_{LOQC}&=&I\nonumber\\
&=& n-S(\rho_{AB})
\label{eq:wloqc}
\eeqa
since Alice can just send her part of the state to Bob, who can
then perform local operations on it to draw all the information.
The quantity
\beq
\Delta_q \equiv I_{LOQC}-\W \s .
\label{eq:deltaq}
\eeq
then tells us how much more information can be obtained when the
channel is changed from a classical channel to a quantum channel.
It is the {\it quantum deficit}.

It is easy to verify that the classical deficit $\Delta_c$ plus
the quantum deficit $\Delta_q$ are equal to the total amount
of correlations contained in the state.
I.e.
\beq
I_M=\Delta_q+\Delta_c
\eeq


Remarkably, for pure states, 
it was found that $\Delta_q=E_D$ where $E_D$ is the
amount of distillable entanglement contained in the state (i.e. the number of
singlets per state $\rab$ that can be drawn under LOCC from a large number of
copies of $\rab$) \cite{JHHH}. 
This was also conjectured to be true for sets of states such as the "maximally correlated" state
or \cite{Rains}.
In general, the quantum deficit $\Delta_q$ can
be due to entanglement, as in the case of pure states, but it also appears that
separable states can have a non-zero $\Delta_q$ as in the case of mixtures
of states which are separable, but indistinguishable \cite{sausage}.  States
such as Werner states \cite{werner} are believed to have $\Delta_q>E_D$.
However, it is not yet clear whether $\Delta_q>E_D$ in the case of many copies.

Under LOCC, $\W$ is the amount of classical information (pure separable states) that can be
extracted using the state $\r$, and used to perform physical work.
$E_D$ has the interpretation
of the maximal amount of useful {quantum logical work} which can be extracted from the
state $\rab$
(since each singlet can be then used for such tasks as the
teleportation of one qubit, or double dense coding).
Here, we will take quantum work to mean teleportation of qubits,
but it is certainly not excluded that there are other forms of quantum work 
\cite{f4}.

Rewriting Equation (\ref{eq:deltaq}) therefore divides the total informational content
into the classical information $\W$ which can be used to perform classical tasks such
as physical work, and the quantum informational part $\D$ which for pure states is equal
to the amount of entanglement which can be used to perform teleportations.
\beq
I=\W+\D
\label{eq:division}
\eeq

However, in
the next section, we will see that the amount of extractable information is {\it not}
$\W+\D$ but rather $\W$ {\it or} $\D$.

%

\section{Complementarity between classical and quantum information}
\label{sec:comp}
We will now show that classical information
and quantum information are complimentary  --
one can use the classical information, or the quantum information,
but not both.
Before discussing the general case, it may be useful
to first show how this complementarity principle plays out with
a simple state such as the singlet of Equation (\ref{eq:singlet}).
This will be done in Subsection \ref{sec:workcomp}, and we will also
show how this can be extended to other states.
In Subsection \ref{sec:infocomp} we will discuss this
complementarity in more generality, and show two different and useful ways
that it can be expressed.
Furthermore, for pure states, one can express the relationship in a particularly simple form,
where the entanglement plays the role of Planck's constant $\hbar$.
In Subsection
\ref{sec:compare} we show that these complementarities are of
the same form as the more familiar
ones encountered in quantum mechanics between conjugate observables such as
position and momentum.
Finally, in Subsection \ref{sec:pure} we give an example of information extraction
which illustrates
our complementarity principle, and shows that pure states can be thought
of as the counterpart to coherent states (i.e. minimum uncertainty
wavepackets).
\subsection{An example of complementarity between quantum and classical information}
\label{sec:workcomp}

Although the mutual information of the singlet is two bits,
initially, neither Alice nor Bob can obtain any information, 
since their local density matrices
are maximally mixed.
However,
in \cite{JHHH} we showed that one bit of information can be obtained by
the two parties.  This can be done
using the following process which was proven to
be optimal.

(a) Alice
uses a measuring device represented by a qubit prepared in the standard
state $|0\ra$ \cite{f5}. She performs a 
cnot \cite{f6} 
using her original state as the control
qubit, and the measuring qubit as the target.  (b) The measurement qubit
is now in the same state as her original bit and can be dephased (i.e.
decohered) in the $|0\ra,|1\ra$ basis so that the information is purely
"classical" (dephasing simply brings the off-diagonal elements of the density
matrix to zero, destroying all quantum coherence).
It is worth noting, that during the dephasing process, one bit of information
is irreversibly transfered into the environment and is no longer available.
(c) The measuring qubit can now be sent to Bob who (d) performs a cnot using the
measuring qubit as the control. His original qubit is now in the standard state
$|0\ra$.
(e) Bob sends the measuring qubit back to Alice who (f) resets the measuring
device by performing a cnot using her original bit as the control.  Alice's
state is now maximally mixed, while Bob's state is known.  They have obtained
one bit of information.  This information can be
used to extract one bit of physical work from a heat bath using a Szilard heat engine.

This process, though optimal, only extracts one bit of classical information, even though two bits of information
could be extracted by someone who is not constrained by LOCC.
However, the singlet also has one bit of quantum information,
which can be used to teleport a single qubit.  In this case, the ability to obtain classical information will
be lost.  If
Alice wishes to teleport the state $\psi_{A'}$ using a singlet, the total initial
state is \be
\psi_{A'} \otimes \psi^-_{AB}
\ee
The final state (after Alice resets her measuring device)
is
\be
{1\over 4}I_{A'A} \otimes \psi_B
\ee
Thus the state (excluding the teleported state $\psi_{A'}$) is now maximally
mixed, containing no classical information. 
One might  think that there could be some other, more sophisticated
protocol that allows one to teleport a qubit in such a way that
the final state will not be maximally mixed. However this is not the case.
All perfect fidelity teleportation schemes were considered by Werner
\cite{werner} who showed that essentially the standard  teleportation
protocol is unique.

We therefore see that for the singlet, there appears to be a complementarity
between teleportation, and classical information - one must choose which one to
obtain, and one bit of information gets destroyed.
The example of the singlet leads to the following general procedure
and result.

For a given state, one can use a particular process  to
distill singlets (quantum information) and perform quantum logical
work. For this process, the amount of singlets
need not be optimal (i.e. less than $E_D$).  Similarly, the classical
correlations can be exploited to obtain classical information under a process which
need not be optimal (i.e. less than $\W$).  After distilling singlets from a
state, one can then use the rest of the state to gain classical information and visa
versa. One can also consider more general processes $\pcal$ where both classical
and quantum information is extracted.  We will denote by $E_D(\pcal)$ and $\W(\pcal)$
the amount of quantum and classical information that can be extracted under this
process.


We will now show that for any LOCC process $\pcal$
\beq
E_D(\pcal) + \W(\pcal) \leq \W
\label{eq:comp1}
\eeq

I.e. the amount of classical information plus quantum information which can be drawn
from a state under any process cannot exceed the optimal amount of
classical information that can be drawn under LOCC.  To see this, we will demonstrate
that if there is a process $\pcal$ such that the bound of Equation
(\ref{eq:comp1}) is violated, then there must exist a process $\pcal'$
which would enable us to draw a greater amount of classical work than the
optimal amount $\W$.
The process $\pcal'$ is as follows:  We first apply the process $\pcal$
to draw an amount $\W(\pcal)$ of quantum information (pure separable states), and $E_D(\pcal)$ of
singlets from the state $\rho_{AB}$ (we don't perform teleportations yet).

Alice and Bob can then obtain more classical information
by converting the $E_D(\pcal)$ singlets into $E_D(\pcal)$ pure separable states
using the optimal procedure described above to convert each
singlet into one local pure state.
Using this process, Alice and Bob, can draw $\W(\pcal)+E_D(\pcal)$ bits
of classical information from the state $\rab$.  Since $\W$ is the optimal amount
of information, the bound given by Equation (\ref{eq:comp1}) follows.

Eq. (\ref{eq:comp1}) shows that there is
a {\it trade-off} between two different processes: if we define the
goal (i) as having one bit of classical information on either site
and goal (ii) as sending one bit of quantum information
from one site to another, then only one of the goals can
be reached. This represents  the trade-off. However, there is more
going on here than a trade-off.  Namely reaching (i)
we {\it irreversibly destroy} the possibility of access to (ii) and vice versa.
This is what corresponds to {\it complementarity}.
All this can be seen easily in the scenario  before the teleportation
process: Alice and Bob share a singlet and Alice has an unknown qubit.
The latter does not change the balance because,
as an additional  resource, it must be counted in both the input and the output.
Then to achieve (i) we can only spend the singlet which
can finally lead to one  classical bit according
to the result of Ref. \cite{JHHH}. This  {\it destroys} all the quantum correlations
and consequently the possibility  to reach goal (ii). If,
on the other hand Alice and Bob decide to teleport,
then goal (ii) is reached but finally Alice's state is completely mixed
(Bob's qubit is in an {\it unknown} pure qubit that does not enter the balance), so classical information
has been  irreversibly  destroyed to enable us to obtain (i).

It is worthwhile to compare Equation (\ref{eq:comp1}) with Equation (\ref{eq:division})
since in a number of cases \cite{JHHH}, $\D=E_D$.  In this case,
Equation (\ref{eq:division}) gives $E_D+\W\leq I$.  One can see that the optimal
amount of distillable quantum information plus the amount of distillable classical information is
in general much greater than the amount that is actually extracted because of the
complementarity between the two.  There is an irreversible process which
destroys our ability to obtain one kind of information, if the other kind
is obtained.

\subsection{Complementarity between quantum and classical information
expressed in terms of entropies}
\label{sec:infocomp}

Although
Equation (\ref{eq:comp1}) essentially expresses the complementarity between
classical and quantum information, however, it is not in a form
usually associated with uncertainty relations.  We will therefore re-express
our bound in two different ways.  First, we will rewrite it as an informational
bound, where the right hand side is a constant, as opposed to something
which depends on the state.  We will also re-express it as a bound
on entropies.  In this case,
the right hands side is related to the entanglement of a state.
Both these bounds have a form like those associated with
complimentary observables.

Let us first re-express (\ref{eq:comp1}) so that the right hand side is
independent of the particular state chosen.  To do this, we write it in terms of the classical information
which is extractable from correlations (as opposed to the local informational
content).  Defining the information which can be extracted from correlations
as $\Icor\equiv\W-\ILO$, we can subtract $\ILO$ from both sides of (\ref{eq:comp1})
and use Equation (\ref{eq:deltac}) to give
\beq
E_D(\pcal,\rab) + \Icor \leq \Dc
\label{leqDc}
\eeq


Under some assumptions in \cite{JHHH} we have proved that $I_{c}\leq n - S_{X}$, $X = A,B$. 
We believe, that this is true in general. Since $\Dc = n - S_{A} - S_{B} - I_{c}$,
we than would obtain $ \Dc \leq min(S_{A},S_{B}) \leq \log d $, so that (\ref{leqDc})
would take the form

\beq
E_D(\pcal,\rab) + \Icor \leq \log d
\label{eq:constantrhs}
\eeq
This bound is the tightest bound one can have which is state independent,
as it is saturated by maximally entangled states, since for two qubit states
we have $E_D(\pcal,\rab) + \Icor \leq 1$.

We can also re-express the complementarity
relation  in terms of
entropies, which will also be useful in relating our complementarity
to the ones usually encountered in quantum mechanics.
We therefore rewrite Eq. (\ref{eq:comp1}) in the form
\beq
\HP + \HL \geq n+ E_f -\W 
\label{eq:comp2}
\eeq
where $\HP$ is defined, in analogy with Eq. \ref{eq:workinfo}
through
\beq
\WP \equiv n-\HP
\label{eq:physical}
\s .
\eeq
The quantity $\HP$ can be thought of as
the Shannon entropy as Alice and Bob would perceive it during the classical information localizing
procedure. In fact, it has been advocated \cite{unruh} that entropy should always
be defined with respect to ones measuring apparatus and how they can be used.
For example, the coarse-grained entropy is defined with respect to 
detectors which can only probe with a finite resolution.  Here, the measuring
devices of Alice and Bob, are restricted to LOCC operations.

Also in analogy with Eq. (\ref{eq:workinfo}), $\HL$ is defined through
\beq
\WL \equiv E_f-\HL
\label{eq:logical}
\eeq
Instead of $n$ which is the number of qubits needed to create the state
$\rab$, one ought to define $\HL$ with respect to $E_f$ -
the number of singlets needed to create the state under LOCC (called the
entanglement of formation). The definition of $\HL$ simply reflects the fact
that not all the entanglement can be distilled to perform teleportations -- there is
''bound entanglement'' \cite{bound}.  Here, since the process is not necessarily
optimal, $\HL$ can be less than the bound entanglement.
The relationship between bound entanglement and entropy (or
heat) was discussed in \cite{thermhor}.

Our definitions help elucidate the strong parallels between entanglement
and classical information.  $n$ separable pure states enable one to perform $n$ bits of
physical work, while $E_f$ singlets allow one to perform $E_f$ bits of quantum
work such as teleportation.  To create a state $\rab$ Alice and Bob will also
need to use $n$ pure separable states, but they will also need $E_f$ singlets.
The entropy $\HP$ prevents Alice and Bob from extracting the full $n$ bits of
classical information, while the bound entanglement $\HL$ prevents them from extracting
the full $E_f$ bits of quantum information.

The information-theoretic version of our complementarity relation takes
a particularly simple form for pure states.
For pure states, it was shown in \cite{JHHH} that $\W=n-E_D=n-E_f$.
We therefore have
\beq
\HP(\psi) + \HL(\psi) \geq 2E_D(\psi) \s .
\label{eq:compure}
\eeq
%
\subsection{complementarity under LOCC compared with ordinary complementarity}
\label{sec:compare}

Although the relations given above, may seem unfamiliar, they actually have a logical structure
similar to the usual
complementarity principle
between non-commuting observables such as $x$ and $p$.  

The reason that Eqs. (\ref{eq:comp1}), (\ref{eq:comp2}) and (\ref{eq:compure})
don't immediately strike one as being like the usual complementarity
relationship, is because we are used to seeing them written like
a Heisenberg uncertainty principle such as
\beq
\Delta x\Delta p \geq \hbar
\eeq
or for general operators $M$, $N$, the Robertson inequality \cite{robertson}
\beq
\Delta M \Delta N \geq \la \psi [M,N] \psi \ra \s .
\label{eq:robertson}
\eeq

However, it is now recognized that these inequalities can
be better expressed as relationships between entropies.
This method of writing the uncertainty principle was begun
by Bialynicki-Birula and Mycielski \cite{birula}, and later advocated
by Deutsch \cite{deutsch} who was dissatisfied with the fact that the bound
on the right hand side of Eq. (\ref{eq:robertson}) is not a constant
but instead depends on the state.  His bound was improved by
Partovi \cite{partovi}, Kraus \cite{kraus}, and Maassen and Uffink \cite{uffink}.
The latter bound can be written as
\beq
H_{\hat{M}}(\psi)+H_{\hat{N}}(\psi)\geq -
2\ln{(sup|\la m|n \ra|)}
\label{eq:uffink}
\eeq
where $m$ and $n$ are the eigenstates of two operators
$\hat{M}$, $\hat{N}$ and the entropies $H_{\hat{M}}$ and $H_{\hat{N}}$ of the
state $\psi$ are the usual Shannon entropies defined for example by
\beq
H_{\hat{M}}(\psi)=-\sum_m|\la m | \psi\ra|^2 \ln{|\la m | \psi\ra|^2} \s .
\eeq

That an uncertainty principle can be written in such a form
makes intuitive sense, because having a larger Shannon entropy
in a certain basis corresponds to a larger uncertainty in measurements
which correspond to that basis.
We therefore see that our complementarity principle is closely related
to the more familiar one encountered in quantum mechanics.

For position and momentum, the Partovi bound takes the form
\beq
H_x(\psi)+H_p(\psi)\leq 2 \ln[2/(1+\delta x\delta p/2\pi\hbar)]
\eeq
for small values of $\delta x\delta p/2\pi\hbar$, where $\delta x$ and
$\delta p$ are the resolution of the detector (i.e. phase space is divided into cells).

Comparing this to Equation (\ref{eq:compure}) we see that for
pure states, the entanglement plays a role analogous to Planck's constant $\hbar$.
The difference of course is that $\hbar$ is a constant which is
independent of the state.  In our case, fixing
the amount of entanglement
in the allowable states is equivalent to fixing $\hbar$ and the detector
resolution.  The right hand
side only depends on the amount of entanglement of the state, and not
on any other properties.
%
It is the addition of entanglement into the system which
acts like $\hbar$ and creates this complementarity.

Our informational complementarity principle, expressed by Equation (\ref{eq:constantrhs})
does have the appealing feature that the right hand side is completely independent
of the state.  It has the form of the informational bound
derived by Hall \cite{hall} for complimentary observables, which is given by
\beq
I_{\hat{M}}+I_{\hat{N}}\leq \log N
\eeq
where $I_{\hat{M}}$ and $I_{\hat{N}}$ gives the amount of
information obtainable from a measurement of complimentary observables $\hat{M}$ and $\hat{M}$,
and  $N$ is the dimension of the Hilbert space.
The similarity between this equation, and Equation (\ref{eq:constantrhs}) is striking.


\subsection{Example: drawing quantum and classical information from pure states}
\label{sec:pure}

We will now consider a protocol $\pcal$ on pure state where both quantum and
classical information is extracted optimally.
It will be used to show the balance between classical and quantum information.
We will also see that pure states can be thought of as being analogous
to coherent states.

Essentially, the procedure is that Alice will perform a measurement which
determines how much entanglement is available.  Depending on the result
of the measurement,
the parties can choose whether they want to extract quantum information,
or classical information.  For example, they may choose to
extract quantum information when they find a lot of entanglement
(i.e. more than the average optimal amount $E_D$)
and extract classical information when there is a small amount of
entanglement (since in this case, they can extract more
classical information than the optimal amount $\W$.


The scenario is similar to the concentration of entanglement scheme of Ref.
\cite{BBPS}. Alice and Bob share n pairs of a pure state
$ \psi_{AB} = a|00\ra + b|11\ra $.
Alice performs a measurement with $n+1$ outcomes. As a result  Alice
and Bob share a maximally entangled state with Schmidt rank
$ d_{k} = \left({n \atop k}\right)$
with probability
$ p_{k} = \left({n \atop k}\right)a^{2k}b^{2(n-k)}, \quad k = 0,...,n $.
The singlet is ''diluted'' into all 2n qubits. 
However, it is not a maximally entangled state of those $2n$ qubits,
so that it can be swapped into a smaller number $\log d_{k}$ of
qubit pairs. Then the remaining pairs will be in product states.

Each process $ \rho \to \{ p_{k}, \rho_{k}\} $
after which information $I_{k}$ is extracted from $\rho_{k}$ with
probability $p_{k}$,
provides $\W = \sum_{k}p_{k}I_{k} - H(\{p\}) $ of information.
The Shannon entropy $H(\{p\})$ of distribution $\{p_{k}\}$
equals the cost of erasure of information which allows
Alice and Bob to work with an ensemble of $\rho_{k} $'s \cite{landauer}.
Thus in our example Alice and Bob have to put $ I_{er} = H(\{p\}) $
of erasure to pay for the next part of the scheme, in which they  draw
the $\sum_kp_kI_{k}$ amount of information.


In our protocol, Alice and Bob will decide whether to extract
entanglement or classical information based on the result of
Alice's measurement (i.e. what value of $k$ she measures).
They divide the outcomes of $k$ into two sets, $K_q$ and $K_c$.

If they obtain outcome $ k \epsilon K_q$ they
i) concentrate the ``diluted'' singlet, obtaining on average
\be
E_D(\pcal)=\sum_{k\epsilon K_q} p_{k}\log d_{k}
\nonumber
\ee
singlets

ii) draw classical information from the rest of qubits, obtaining on average
\be
I_{c_{1}}(\pcal) =\sum_{k\epsilon K_q} p_{k}(2n - 2\log d_{k})
\ee
If instead they obtained the outcomes with $k\epsilon K_c $ they

iii) draw classical information directly from the state with the average result
\be
I_{c_{2}}(\pcal) =\sum_{k\epsilon K_c} p_{k}(2n - \log d_{k}) \s .
\ee

Summing up all information drawn from the system, we have
\be
I(\pcal) = E_D({\pcal}) + I_{c_{1}}(\pcal) + I_{c_{2}}(\pcal) - I_{er},
\ee
which gives
\be
I(\pcal) = 2n - \sum_{k = 0}^{n}p_{k}\log d_{k} - H(p).
\label{bilans}
\ee
Passing to intensive quantities, $\tilde i\equiv{I(\pcal)\over n}$
is asymptotically equal to the maximum possible amount of  classical information
per pair which can be obtained starting with
${|\psi_{AB}\ra\la\psi_{AB}|}^{\otimes n}$,
namely $\tilde i \approx 2 - S_A$, where $S_A$ is the entropy of reduction of $\psi_{AB}$.
Indeed the last term in (\ref{bilans}) is of order of $\log n$,
so that its contribution vanishes in the asymptotic limit.
Thus the inequality (\ref{eq:comp1}) of complementarity
is saturated.  It follows that for pure states it is possible to
obtain partially quantum and partially classical information, without
any loss.  However for mixed states it is rather unlikely that an
optimal protocol  which saturates the inequality would exist.

Since the bound is saturated, and Alice and Bob may obtain any
amount of either $\W(\pcal)$ or $E_D(\pcal)$  (up to their maximal value),
we can therefore think of pure states as being analogous to coherent states i.e.
minimally uncertain states.
Maximally entangled states are also the only ones which also saturate
the constant bound of (\ref{eq:constantrhs}).

\section{LOCC commutator}
\label{sec:loccom}
In the previous sections, we introduced a complementarity principle
between quantum and classical information.  Physically,
it referred to general processes, rather than any particular implementation.
It would therefore be useful to see if there is a complementarity principle
which just refers to
general measurements or operations.  Indeed, we will find that when the
implementation of a measurement is restricted to LOCC, it induces a new
set of complementarities.  One can generalize this further and consider
complementarities when one is restricted to other classes of operations.

We shall start with two examples.  In Subsection \ref{sec:parityphase} we
will demonstrate that the parity and phase operator, which normally commutes
in quantum mechanics, no longer  commutes under LOCC.  In
subsection \ref{sec:sausage} we will look at measurements which distinguish
between the orthogonal states of reference \cite{sausage}.
Finally, in Subsection \ref{sec:commutator} we will introduce the notion
of the restricted commutator
that is induced between two observables,
if our means to measure them are somehow restricted. Our definition will hold
for any class of operations, but for clarity we will talk about LOCC operations.
\subsection{The parity and phase observable}
\label{sec:parityphase}
Consider two observables $\Sx=\sxa\otimes \sxb$ and
$\Sz=\sza\otimes \szb$, where $\sigma_x$ and $\sigma_z$ are the usual
Pauli spin matrices and the superscript refers to which subsystem
it acts upon. They
commute,
\be
[\Sx,\Sz]=0
\ee
and their eigenbasis is the Bell basis
\beqa
\psi^\pm&=&|01 \ra \pm |10 \ra \nonumber\\
\Phi^\pm&=&|00 \ra \pm |11 \ra
\eeqa

$\Sz$ measures the parity bit, and will therefore distinguish between the
$\psi$ and $\Phi$ eigenstates, while
$\Sx$ measures the phase bit, and will distinguish between the eigenstates
which have a $+$ as the relative phase, or a $-$.
E.g, if one finds $\Sz=0$ and $\Sx=0$, then
we have a singlet.

However if Alice and Bob are restricted to LOCC
operations and want to measure such observables on their shared system, it is
impossible. Indeed, to measure $\Sx$, Alice and Bob must separately measure
$\sigma_x$, while to measure $\Sz$ they have to measure separately $\sigma_z$.
Clearly, since $\sigma_x$ does not commute with $\sigma_z$, they cannot measure
both the parity and the phase.
One might suspect, that there could exist some complicated LOCC protocol that
measures them jointly, somehow avoiding measuring directly local non-commuting
observables. Later we will show by a simple argument that this is impossible in
general by any LOCC operation. However, here we would like to grasp the rough
idea of the difference between the global and LOCC measurement.

To this end, note, that in the distant labs case, Alice and Bob measure too
much. Indeed, measuring $\Sx$ globally,
gives one bit of information (phase bit), because $\Sx$ has only two
eigenvalues. In contrast, measuring the phase locally (by having Alice
and Bob measure $\sigma_x$ on their subsystem), the two parties
will acquire 2 bits of information.
Thus in local measurements the measurement is nondegenerate, while
$\Sx$ and $\Sz$ are degenerate.  In fact, a local determination of parity
and phase {\it must} acquire two bits of information.


We can simulate a global measurement of parity or phase by
using the local operators
%
%
\be
\Szl=\sigma_z^{(A)} + \alpha_z \sigma_z^{(B)}
\ee
\be
\Sxl=\sigma_x^{(A)} + \alpha_x \sigma_x^{(B)}
\ee

where the parameters $\alpha$ act to break the degeneracy in the operators
$\Sx$ and $\Sz$. Then these local measurements of parity and phase no longer
commute
\be
[\Sxl,\Szl]= -i(
\sigma_y^{(A)} + \alpha \sigma_y^{(B)}
)
\label{commutator}
\ee
where we have redefined the constants $\alpha$.
Thus this measurement of parity and phase cannot be jointly measured under LOCC.

%
%

Here, we have only given one possible local realization of the parity
and phase measurement.  It therefore may be possible that one can
find a clever procedure, perhaps involving positive operator valued
measured (POVM's), such that the parity and phase measurements commute.
This, however, cannot be the case.

Let us imagine that there exists local implementations of $\Sz$
and $\Sx$ which are jointly measurable (i.e. commute).
Then we would be able to use these locally implementable operators
to
distinguish between the four Bell states. This is however impossible,
as shown in \cite{aditi}.   Naively, this is because
if one could distinguish between the Bell states, then one can
produce entanglement from the identity state (which is separable).
This would contradict the fact that entanglement cannot be created under LOCC.
In fact the
problem is more subtle. One could, in principle
be able to distinguish the Bell states but in so doing, the
entanglement could be destroyed. Indeed, in the case of two entangled states,
one can distinguish them by LOCC \cite{twoent}
For this case, the entanglement
is necessarily destroyed during the measurement. However,
distinguishing between the four Bell states
would lead to entanglement creation under LOCC \cite{aditi}.  We therefore see that
that the parity and phase cannot be jointly measurable.  In fact,
parity and phase must be completely complimentary, since if one was
able to measure the parity, and get even partial knowledge of the phase,
then one would be able to create entanglement.

It is also interesting to ask how much entanglement is needed in order
to implement the parity and phase operators in such a way that they
commute.  The answer, is one bit of entanglement.  To see this,
we note that 1 bit of entanglement is clearly sufficient, since
Alice can use a singlet to teleport her qubit to Bob, who can
then measure the parity and phase.  1 bit of entanglement must also
be necessary, since measuring parity and phase under LOCC allows one to create
1 bit of entanglement \cite{aditi}.  Since one cannot create entanglement
under LOCC, one must use up at least 1 bit of entanglement to make the
measurement. It is therefore rather interesting that if we act the commutator of
Eq. (\ref{commutator}) on a separable state then we can get an entangled state,
which is maximally entangled for $\alpha = 1$.

%

Finally, it is worth asking whether one can find other examples for
$2X2$ systems.  In other words, are there other observables which commute
globally, but do not commute under LOCC. It appears that the number of examples
is very limited. For pairs of product observables of the form $A\otimes B$ there
is (up to local unitary transformations) only one other pair of operators which
commute globally, namely
$\sxa\otimes\syb$ and $\Szl$.

The proof of this result is contained in the appendix to this paper.  For now,
we simply state the result:

{\it {\bf Proposition} -
  If for some products of two qubit observables
  $[A \otimes B, C \otimes D]=0$
  then up to unitary product transformation
  $U_{1} \otimes U_{2}$ and constant factor, one has
  \begin{equation}
    A=B=\hat \sigma_{z},
  \end{equation}
  \begin{equation}
    C=D=\hat \sigma_{x}.
  \end{equation}
  }
\subsection{distinguishing separable states}
\label{sec:sausage}
In \cite{sausage}, a set of nine states which are orthogonal and separable
were presented.
It was then proven, that although they are orthogonal, they cannot be
distinguished under LOCC.  These states (often referred to as "sausage states")
therefore exhibit a form of non-locality without entanglement.  One can however
distinguish between some of the states.  There are therefore operators which can
be constructed using the sausage states as basis states.  As an example, we will
construct two such operators which although they commute globally, and
are implementable locally, do not commute under LOCC.

The nine sausage states are (using a different numbering scheme from
 \cite{sausage} for convenience)

\begin{equation}
  \begin{array}{lll}
    \,&{\mbox \small A}\ &
    {\mbox \small B}\\
    \psi_1=&|0+1\rangle&|2\rangle\\
    \psi_2=&|0-1\rangle&|2\rangle\\
    \psi_3=&|0\rangle&|0+1\rangle\\
    \psi_4=&|0\rangle&|0-1\rangle\\
    \psi_5=&|1+2\rangle&|0\rangle\\
    \psi_6=&|1-2\rangle&|0\rangle\\
    \psi_7=&|1\rangle&|1\rangle\\
    \psi_8=&|2\rangle&|1+2\rangle\\
    \psi_9=&|2\rangle&|1-2\rangle \s .\\
  \end{array}
  \label{9states}
\end{equation}

Now we can construct an operator $O_1$ which has as it's eigenstates, the
first seven states with seven different, non-zero eigenvalues and remaining two
eigenstates with zero eigenvalues. We can also construct an operator $O_2$ which has 
$\psi_8$ and $\psi_9$ as eigenstates with two different, non-zero eigenvalues and 
remaining seven eigenstates with zero eigenvalues. 
These operators clearly commute globally, since all the $\psi_i$
are orthogonal. However, under LOCC they clearly cannot commute.  If they did,
one could measure $O_1$ and $O_2$ simultaneously, and therefore, distinguish
between all nine sausage states, in violation of the indistinguishability proof
given in \cite{sausage}.

Now, $O_2$ can easily be implemented under LOCC.  Bob can simply use the
projectors $|1+2\ra$ and $|1-2\ra$ to implement $O_2$.  These projectors
distinguish between $\psi_8$ and $\psi_9$.
$O_1$ can also be implemented under LOCC, although some effort is needed.
Consider for example an implementation $O_1'$ which
instead has the following orthogonal eigenbasis.
\begin{equation}
  \begin{array}{lll}
    \,&{\mbox \small A}\ &
    {\mbox \small B}\\
    \psi_1=&|0+1\rangle&|2\rangle\\
    \psi_2=&|0-1\rangle&|2\rangle\\
    \psi_3=&|0\rangle&|0+1\rangle\\
    \psi_4=&|0\rangle&|0-1\rangle\\
    \psi_5=&|1+2\rangle&|0\rangle\\
    \psi_6=&|1-2\rangle&|0\rangle\\
    \psi_7=&|1\rangle&|1\rangle\\
    \psi_{10}=&|2\rangle&|2\rangle\\
    \psi_{11}=&|2\rangle&|1\rangle\\
  \end{array}
  \label{9modifiedstates}
\end{equation}

The first seven eigenstates are identical to the eigenstates of $O_1$,
and so $O'_1$ is an implementation of $O_1$.  Furthermore,
$O'_1$ can be implemented under LOCC using a sequence of
von Neumann projection measurements which was given in \cite{sausage}.
The detailed procedure is contained in Appendix \ref{ap:sausage}.

The difficulty, is  that while the eigenvectors $\psi_1$-$\psi_7$
commute with $O'_2$, the projectors onto  $\psi_{10}$ and $\psi_{11}$ do not.
If we write $O'_1$ and $O'_2$ as
\be
O'_1=O_1+|2\ra\la 2|\otimes\szb, \s\s O'_2=|2\ra\la 2|\otimes\sxb
\ee
where we once again use the Pauli matrices, this time written in the $|1\ra$,
$|2\ra$ basis, then we find that while $[O_1,O_2]=0$, we have
\be
[O'_1,O'_2]=-i|2\ra\la 2| \otimes \syb
\ee

Unlike the case of the parity-phase commutator, this commutator, operating
on a separable state, cannot create entanglement.  This may be
related to the fact that for parity and phase, the eigenstates
are entangled, while for $O_1$ and $O_2$, the eigenstates are separable.
\subsection{The LOCC commutator}
\label{sec:commutator}

In general, we may define the LOCC commutator as follows.  Consider
two operators $\M$ and $\N$ which are implementable under LOCC.
For all LOCC implementations $\Ml$ and $\Nl$ we can define the LOCC commutator
as
\be
[\M,\N]_{LOCC}=\min{[\Ml,\Nl]}
\ee
where the minimum is taken over all LOCC implementations of $\M$ and $\N$.
It is not clear what the physical significance is of the right hand
side of this equation.  In the parity-phase it was an operator which could
create one bit of entanglement (which was also the amount of entanglement
needed to implement both measurements).  This is perhaps intriguing in light of
Equation (\ref{eq:compure}) where entanglement played the role of $\hbar$.

The LOCC commutator can also be generalized to any class of operations.  If we
consider the set of all allowable operations $A$, and a restricted subset of
these $R\subseteq A$, then we can define a restricted commutator much in the
way we have done here.
%

There is however, one key difference between the interpretation of this
commutator, and the usual commutator we are familiar with.  The fact
that two operators $\M$ and $\N$ do not LOCC commute can imply that
they are complementarity observables.  It will also imply that
one cannot prepare a state under LOCC which has a definite value of the
observable $\M$ and $\N$.  But it does not imply uncertainty of measurements.

To see this, recall that a singlet (for example) has definite parity and phase.
If Alice and Bob are given an ensemble of singlets, and measure the phase on half
the ensemble they will always get the same result ($-$).  If they measure
the  parity on the other half of the ensemble, then they will also
always get a definite result (0).  
There is nothing uncertain about what the outcome of a measurement of
parity {\it or} phase will be. 
However, if Alice and Bob are given an unknown state (perhaps a
singlet) then their measurement of phase will completely destroy 
their ability to determine what the parity is, and visa versa.
For a single state, they cannot determine both the parity and the phase.
Likewise, there is no way for them to prepare a system with definite
parity and phase (which would amount to creating entanglement).

This decoupling of complementarity and uncertainty shows
that they are independent concepts.  The essential reason for this decoupling
is that the measurement does not prepare the system in an eigenstate
of the observable we are trying to measure.  The measurement irreversibly
alters the state.  Usually, the von Neumman postulate holds -- after a measurement,
the state is in an eigenstate of the observable.  Therefore it was hard to distinguish
between complementarity and uncertainty, and discussions concerning this difference
could seem speculative and philosophical.  However, as we have seen, the situation
changes in the LOCC paradigm.

\section{Conclusion}
\label{sec:conc}

We have found that when one is restricted to making only local measurements
and communicating classically, then new types of complementarities are
induced.  One type of complementarity was between classical and quantum
information, and was given by Equation (\ref{eq:comp1}).  
If one attempts to maximize the amount of classical information 
(pure separable states) then the ability to extract quantum information and perform quantum operations such as
teleportations is lost.  Likewise, extracting quantum information
destroys the ability to obtain classical information.  

In Equations (\ref{eq:constantrhs}) and (\ref{eq:comp2}),
we wrote this complementarity
in a form which was of the same kind as those ordinarily
encountered in quantum mechanics.  For pure states, we found
that the entanglement plays the role of $\hbar$.

We also found a complementarity that gets induced between operators when implemented under LOCC.  For
example, the parity and the phase of a $2X2$ state is no longer jointly
measurable.  It is remarkable that in this case, the uncertainty principle and complementarity are decoupled,
and one can have complementarity without uncertainty.  We therefore see that they are indeed
separate concepts\cite{Rosjanie}.

We then introduced the notion of the LOCC commutator, to quantify
this complementarity.  How to interpret this quantity is an interesting open question.
In this regard, it is perhaps interesting that for parity and phase, the
LOCC commutator can create one bit of entanglement.  Likewise, being able
to measure parity and phase simultaneously also results in the creation
of one bit of entanglement.

It therefore might be interesting to ask how much entanglement would be needed
such that one can jointly measure two observables.  Quantifying this
"entanglement assisted commutator" might help answer some of the questions
raised here.

It also would be interesting to relate the complementarity principle between
operators, and between classical and quantum information.  The latter involves
comparisons between two types of restricted operations (LO and LOCC), while
the complementarity principle between operators only involves LOCC.  However,
both seem to involve the notion of entanglement.

In this paper, the part of quantum information which was discussed
was entanglement.  However, the quantum deficit $\D$
is also non-zero for unentangled states (at least for a finite number
of copies).  It would therefore be of interest to also consider
the case of date hiding \cite{hiding}, where Alice and Bob are essentially
unable to obtain the classical information encoded in a state.

Finally, the above results support the view that quantum states carry two complementary
kinds of information, the classical information which is locally accessible and quantum
information which can be used for such tasks as teleportation (see in this context \cite{balance}).
This complementarity lies at the foundations of quantum mechanic more deeply 
than it might seem. We believe that complementarity in general is a fundamental and
intrinsic feature of information carried by physical systems which can not be 
derived from any probabilistic models.
\appendix
\section{Proof of Proposition 1}
\label{ap:prop1}

Let us provide a simple lemma first:

{\it {\bf Lemma} - If $X \otimes Y=R \otimes S$
  for some operators $X, Y, R, S$ then
  $X=\alpha R$, $Y=\alpha^{-1}S$
  for some nonzero number $\alpha$.
  }

Proof of the above lemma is immediate.
Without loss of generality we can consider
$X, Y$ to be of full rank  and utilize their inverses
(otherwise they are pseudoinverses) getting
$I \otimes I =X^{-1}R \otimes Y^{-1}S$.
Comparing the eigenvectors of both sides of the latter
formula gives $Y^{-1}S=\alpha I$, $X^{-1}R=\alpha^{-1} I$, 
concluding the proof of the lemma.

Now we shall provide the simple proof of the following

{\it {\bf Proposition} -
  If for some products of two qubit observables
  $[A \otimes B, C \otimes D]=0$
  then up to unitary product transformations 
  $U_{1} \otimes U_{2}$ and constant factor, one has}
  \begin{equation}
    A=B=\hat \sigma_{z},
  \end{equation}
  \begin{equation}
    C=D=\hat \sigma_{x}.
  \end{equation}

{\it Proof - }
By adding and subtracting
the term $CA\otimes BD$, it is immediate that vanishing of the commutator
from the Proposition is equivalent to
\begin{equation}
  [A,C] \otimes BD=CA \otimes [D,B]
  \label{localkom}
\end{equation}

Applying the lemma to the above we get
\begin{equation}
  [A,C]=\alpha CA, \ \ [D, B]=\alpha BD
  \label{both}
\end{equation}

Now for two qubits we put
$A=aI + \vec{a} \vec{\sigma}$, $B=bI + \vec{b} \vec{\sigma}$,
$C=cI + \vec{d} \vec{\sigma}$, $D=dI + \vec{d} \vec{\sigma}$.
We then perform a simple calculation taking into account the fact that (because
of linear independence of $I$, $\sigma_{x}$, $\sigma_{y}$, $\sigma_{z}$)
absence of $I$ on one side implies the same for the other side.
This gives two equations:
\begin{eqnarray}
  &&
  \alpha[ \vec{a} \vec{\sigma}, \vec{b} \vec{\sigma}]=
  (a \vec{a} + c \vec{c}) \vec{\sigma} + i \vec{a} \times \vec{c} \vec{\sigma} \nonumber \\
  &&
  \alpha[ \vec{b} \vec{\sigma}, \vec{d} \vec{\sigma}]=
  (b \vec{b} + d \vec{d}) \vec{\sigma} + i \vec{b} \times \vec{d} \vec{\sigma} \nonumber \\
\end{eqnarray}
Calculating the LHS for both sides and using the linear independence
of Pauli matrices we get finally:
\begin{eqnarray}
  &&
  a \vec{a} + c \vec{c} + i(1- 2 \alpha) \vec{a} \times \vec{c} =0\nonumber \\
  &&
  b \vec{b} + d \vec{d} + i(1-2\alpha) \vec{b} \times \vec{d}=0 \nonumber \\
  \label{final}
\end{eqnarray}

Now let us observe that
(i) $\vec{a} \times \vec{c} \neq 0$
and hence also
(ii) $\vec{a}\neq0$, $\vec{c}\neq0$.
Indeed
if $\vec{a} \times \vec{c} = 0$
then $[A,C]=0$ and consequently
(see (\ref{localkom})) either $AC=0$ (which leads to the trivial
solution because some operators must completely vanish)
or $AC\neq 0$ but then (again because of (\ref{localkom}))
also $[B,D]=0$ which would be trivial again.

Because of (i) and (ii) the LHS of the first line of (\ref{final})
is a linear combination of three {\it nonzero and linearly independent}
vectors $\vec{a}$, $\vec{c}$, $\vec{a} \times \vec{c}$
so all the coefficients in the combination must vanish giving in particular $a=c=0$.
In a similar way we get $b=d=0$. This simplifies our observables:
$A= \vec{a} \vec{\sigma}$,
$B=\vec{b} \vec{\sigma}$, $C=\vec{d} \vec{\sigma}$, $D=\vec{d} \vec{\sigma}$.
Putting them again into (\ref{both}) we get immediately
\begin{equation}
  ( \vec{a} \times \vec{c} \vec{\sigma}) \otimes
  (\vec{b}\vec{d}I + i\vec{b} \times \vec{d} \vec{\sigma})=
  (\vec{a}\vec{c}I + i\vec{c} \times \vec{a} \vec{\sigma})
  \otimes  (\vec{b} \times \vec{d} \vec{\sigma})
\end{equation}
which
for nonzero $\vec{a} \times \vec{c}$ and $\vec{b} \times \vec{d}$
is satisfied iff $\vec{a}\vec{c}=\vec{b}\vec{d}=0$.
We can put $\vec{a}=\vec{b}=\hat{z}$
since we can always choose such a local basis for Alice
and Bob.  We then have $\vec{c}=\vec{d}=\hat{x}$ (again using our choice of
label for the direction orthogonal to $\hat{z}$).  This concludes the proof
of the Proposition.
\section{Measurement of seven sausage states under LOCC}
\label{ap:sausage}
Here we show how to implement $O_1'$ using a {\it ping-pong} process
between Alice and Bob.
Essentially, the procedure is:

b1) Bob first does a projection on $|2\ra$ and
communicates his result to Alice.

a1) If his result is positive, then Alice can
project onto the three states $|0+1\ra$, $|0-1\ra$ which will
distinguish between $\psi_1$ and $\psi_2$.  However, if
she finds neither $\psi_1$ or $\psi_2$ she will know
that the state is $\psi_{10}$ which is in some sense superfluous
information which she would not get if she was measuring $O_1$
globally.

a1') If Bob's first projection yielded a negative result,
then Alice instead projects onto the $|0\ra$ state and communicates
her result to Bob.

b2) If her projection found the state $|0\ra$ then
Bob projects onto $|0+1\ra$ and $|0-1\ra$ which distinguishes
between $\psi_4$ and $\psi_5$

b2') If her result was negative,
Bob projects onto $0\ra$ and $|1\ra$ and  communicates
the result to Alice.

a2)  Alice can then make the final orthogonal projection, either onto $|1+2\ra$
and $|1-2\ra$, or onto $|1\ra$ and $2\ra$ depending on Bob's result.
This distinguishes between $\psi_5$, $\psi_6$ and $\psi_7$, as desired,
but it also singles out $\psi_{11}$ which is again, surplus information
which is not required to implement $O_1$.


\vspace{1in}

{\bf Acknowledgments}:
This work is supported by EU grant EQUIP, Contract No. IST-1999-11053.
M. H., P. H. and  R.H.  acknowledge support of KBN grant No. 2 P03B 103 16.
M.H. would like to thank Viacheslav Belavkin and Berthold-Georg Englert for discussion on uncertainty
principles.
J.O. would like to thank Jacob Bekenstein for interesting discussions and the Vergelle Institute
for their hospitality while part of this work was completed.  He acknowledges the support of
the Lady Davis Fellowship Trust, and
grant No. 129/00-1 of the Israel Science Foundation.


\begin{thebibliography}{9}
\bibitem{Pauli} ``One may view
the world with the p-eye and one may view it with the q-eye but
if one opens both eyes simultaneously
then one gets crazy'' -- W. Pauli in a letter to W. Heisenberg, 19 October, 1926
\bibitem{Heisen}W. Heisenberg, Z. Physik {\bf 43} 172 (1927)
\bibitem{Bohr} N. Bohr, Nature {\bf 121} 580 (1928), Phys. Rev. {\bf 48} 696 (1935)
\bibitem{JHHH} J. Oppenheim, M. Horodecki, P. Horodecki, R. Horodecki, quant-ph/0112074
\bibitem{f1} Although we will only discuss
  bipartite states, the analysis presented here applies equally well for
  multipartite states.  We also work with states made up of qubits, but one
  could just as well consider more general states.
\bibitem{thomascover} T. Cover and J. Thomas, Elements of Information Theory, John Wiley and Sons, Toronto, 1991
\bibitem{f2} By classical system we mean a $d$ level system, the states of
  which are probability distributions over $d$ items.
\bibitem{Bennett92} C. Bennett, S. Wiesner, Phys. Rev. Lett. {\bf 69} 20 (1992)
\bibitem{BSST-entass} More generally, it was shown that
  the mutual information $I_M$ describes the capacity of noisy channel
  assisted  by entangled states. H. Bennett, P.W. Shor, J.A. Smolin and A.V. Thapliyal
Phys. Rev. Lett. {\bf 83} 3081 (1999).
\bibitem{cerf}N. Cerf, C. Adami, Phys.Rev.Lett. 79 (1997) 5194.
\bibitem{Zurek} W. H. \.Zurek, quant-ph/0011039; H. Ollivier, W. H. \.Zurek
quant-ph/0105072, Phys. Rev. Lett. (in print).

  \bibitem{HV} L. Henderson V. Vedral quant-ph/0105028

  \bibitem{THDL-entpur} B. M. Terhal, M. Horodecki, D.W. Leung, D. P. DiVincenzo, quant-ph/0202044

\bibitem{szilard} L. Szilard, Zeitschrift fur Physik, 1929, 53, 840-856, translated into English
and reprinted in Quantum Theory and Measurement, eds. J.A. Wheeler and W. Zurek, Princeton
University Press, Princeton (1983)
\bibitem{vn} J. von Neumann, Mathematische Grundlagen der Quantenmachanik, Springer, Berlin (1932)
translated into English
and reprinted in Quantum Theory and Measurement, eds. J.A. Wheeler and W. Zurek, Princeton
University Press, Princeton (1983)
\bibitem{p_of_forg} A Recent review on connections between
thermodynamics and classical and quantum correlations is given in
M.B. Plenio V. Vitelli, Contemporary Physics {\bf 42}, 25-60 (2001)
\bibitem{f3} We can think of the classical channel as a dephasing channel
which transforms the state $\r$ into a state $\r'=\sum_i P_i \r P_i$
where $P_i$ are orthogonal projection operators which are chosen by
Alice and Bob.
\bibitem{BBPS} H. Bennett, H. Bernstein, S. Popescu and B. Schumacher
Phys. Rev. A {\bf 53}, 2046 (1996)
\bibitem{landauer} R. Landauer, IBM J. Res. Develop. {\bf 5} 183 (1961); C. H. Bennett, Int. J. Phys. {\bf 21}, 905 (1982)
\bibitem{Rains} E. Rains, Phys. Rev. A {\bf 60}, 179 (1999).
\bibitem{sausage} C. H. Bennett, D. DiVincenzo, Ch. Fuchs, T. Mor, E. Rains, P. Shor,
J. Smolin, W. K. Wootters,  Phys. Rev. A {\bf 59}, 1070 (1999).
\bibitem{werner} R. Werner, All Teleportation and Dense Coding Schemes, quant-ph/0003070
\bibitem{balance}R. Horodecki, M. Horodecki and P. Horodecki Phys. Rev. A {\bf 63} 022310 (2001).
\bibitem{f4} For example, data hiding \cite{hiding} might also be considered to be a form
of quantum work.
\bibitem{f5} To ensure that we are actually extracting information from
the state, and not the measuring device, one needs to reset the measuring device
after the entire procedure.
\bibitem{f6} A cnot, or {\it controlled not} gate, flips the state
of a target qubit if the control qubit is in the $|1\ra$ state,
and does nothing otherwise.
\bibitem{bound}M. Horodecki, P. Horodecki and R. Horodecki Phys. Rev. Lett. {\bf 80}, 5239 (1998).
\bibitem{unruh} This operational approach has been advocated by W. Unruh (private communication), and is also contained
in the work of Janes and Gibbs (E. T. Janes, "The Gibbs Paradox", in Maximum Entropy and Bayesian Methods, eds C. R. Smith, G.
J. Erickson, P. O. Neudorfer, Kluwer Academic Publishers, Dordrecht, Holland (1992))
  \bibitem{thermhor}P. Horodecki, M. Horodecki and R. Horodecki Acta. Phys. Slovaca {\bf 47} 148 (1998).
  \bibitem{robertson} Robertson, Phys. Rev. {\bf 34} 163 (1929)
  \bibitem{birula} I. Bialynicki-Birula, J. Mycielski, Commun. Math. Phys. {\bf 44} 129-132 (1975)
  \bibitem{deutsch} D. Deutsch, Phys. Rev. Lett. {\bf 50} 631 (1983)
  \bibitem{partovi} M.H. Partovi, PRL, {bf 50} 1883 (1983)
  \bibitem{kraus} K. Kraus, Phys. Rev. D {\bf 35}, 3070 (1987)
  \bibitem{uffink} H.Maassen and J. Uffink, PRL {\bf 60} 1103 (1988)
\bibitem{hall} M. Hall, PRL {bf 74} 3307 (1995)
\bibitem{aditi}S. Ghosh, G. Kar, A. Roy, A. Sen (De), U. Sen, quant-ph/0106148
\bibitem{twoent} J. Walgate, A.J. Short, L. Hardy, V. Vedral, PRL {\bf 85} 4972 (2000)
\bibitem{hiding} D. P. DiVincenzo, D. W. Leung, B. M. Terhal, quant-ph/0103098
\bibitem{Rosjanie} It would be interesting to relate our discussion of complementarity
and uncertainty with investigations of Refs. \cite{complunc}
\bibitem{complunc}
I. Kim and G. Mahler, quant-ph/0004056; A.  Klachko and A. Schumowsky quant-ph/0203099.
\end{thebibliography}
\end{document}